\documentclass[aps,prd,twocolumn,groupedaddress,amssymb,eqsecnum,showpacs]{
revtex4}
\usepackage{graphicx}
\usepackage{bm}

\begin{document}

\preprint{CMU-HEP-02-10}

\title{The invisible axion in a Randall-Sundrum universe}

\author{Hael Collins}
\email{hael@cmuhep2.phys.cmu.edu}
\author{R.~Holman}
\email{holman@cmuhep2.phys.cmu.edu}
\affiliation{Department of Physics, 
Carnegie Mellon University, Pittsburgh PA\ \ 15213}

\date{\today}

\begin{abstract}
We study the problem of integrating an invisible axion into the
Randall-Sundrum scenario as an example of how to generate new energy scales
between the extremes of the Planck mass and the electroweak scale.  In this
scenario, the axion corresponds to the phase of complex bulk scalar field.  We
show how to generate an intermediate energy scale by including a third brane
in the scenario.  We discuss the stabilization of this brane in detail to
demonstrate that no additional fine-tunings arise. 
\end{abstract}


\maketitle

\section{Introduction}\label{intro}

One of the features which the Randall-Sundrum scenario \cite{rsa} shares with
other solutions to the hierarchy problem is that it assumes a desert between
the scale of electroweak physics and the scale of gravity.  Although new
phenomena---strong gravity and bulk Kaluza Klein modes---appear above the
electroweak scale, all the physics in this scenario can be expressed in terms
of these two scales.  Such a picture is usually adequate since we have no
direct evidence of phenomena between these energies.  Yet in some cases we may
wish to introduce some new physics whose dynamics occur at an intermediate
scale.  The difficulty in the Randall-Sundrum brane world is to understand how
a scale can naturally arise, surviving in the low energy theory, that is not
either of these two natural extremes.

A specific instance of where such an intermediate scale is needed occurs in
the invisible axion solution to the strong CP problem \cite{strongCP}.  The
vacuum structure of QCD combined with the CP violation in the weak
interactions permits an interaction of the form 
\begin{equation}
\bar\theta {\alpha_s\over 8\pi}\, {\rm Tr} [\epsilon_{\mu\nu\lambda\rho}
F^{\mu\nu}F^{\lambda\rho}] 
\label{thetaterm}
\end{equation}
where $F^{\mu\nu}$ is the QCD field strength.  This interaction violates P and
CP and is highly constrained by measurements of the neutron dipole moment
which require $\bar\theta\le 10^{-9}$.  As a free parameter, $\bar\theta$ must
thus be finely tuned for an acceptable theory.  Peccei and Quinn \cite{pq}
showed that if $\bar\theta$ is promoted to a dynamical field $a(x^\mu)$ which
is the Goldstone boson associated with a spontaneously broken global
$U(1)_{\rm PQ}$ symmetry, then $\bar\theta$ is dynamically driven to zero.
Although this field, the axion, is a Goldstone mode, it does acquire a mass of
the order $\Lambda_{\rm QCD}^2/f$ where $f$ is scale at which the $U(1)_{\rm
PQ}$ breaking occurs.  An $f$ of the order of the electroweak scale produces
too massive an axion for experimental constraints.

An acceptable axion mass does occur when the $U(1)_{\rm PQ}$ breaks at some
high scale $f\gg{\rm TeV}$.  From astrophysical observations, this scale
should lie within the interval \cite{turner,cosmological}
\begin{equation}
10^{10}\, {\rm GeV} \alt f \alt 10^{13}\, {\rm GeV} . 
\label{fconstraint}
\end{equation}
Since the invisible axion models do not address the hierarchy problem, they do
not attempt to explain whether such a scale can arise naturally.

In the Randall-Sundrum scenario, the only natural scales are the bulk Planck
mass, $M_5$, and the AdS curvature, $k$.  Other, exponentially smaller scales
do arise when the physics responsible for them is confined to a region at some
distance from the UV brane.  Although the mass scales for the fields confined
to the IR brane are also of the order $M_5$, when the fields there are
rescaled to remove red-shift factors introduced by the induced metric on the
brane, the apparent mass scales on the IR brane can be naturally of the order
of the electroweak scale.  Goldberger and Wise \cite{gw} showed that the
position of the IR brane relative to the UV brane can be stabilized---and thus
the electroweak-gravity hierarchy---without finely tuning the parameters of
the stabilization mechanism.  The observed Planck mass in low energy, four
dimensional effective theory, determined by $M_4^2\approx M_5^3/k$, remains
large.  

In this article we shall use the invisible axion as a case study of how to
introduce new intermediate scales into the Randall-Sundrum scenario.  For this
purpose it provides an ideal subject---the scale $f$ is experimentally
constrained to not be that associated with either of the branes.  These
constraints arise, moreover, from low energy physics with respect to the
electroweak scale so that bulk effects do not allow us to modify these bounds
as in scenarios with large extra dimensions \cite{ddg}.  We shall see that
adding a further brane in the bulk produces an experimentally reasonable value
for $f$ without any new excessive fine-tunings.

The next section discusses the origin of the scale $f$ associated with the
breaking of $U(1)_{\rm PQ}$ when the axion is the phase of a complex bulk
scalar field.  Section \ref{intbrane} discusses how the scale $f$ arises when
a third brane is added to the bulk and analyzes the problem of stabilizing all
the branes with a single real scalar field.  In Section \ref{bulkpot}, we show
that 
it is more difficult to generate an acceptable value for $f$ using a bulk
potential for the complex field containing the axion.  Section \ref{conclude}
concludes.

\section{The invisible axion as a bulk field}\label{prelude}

The action for the original Randall-Sundrum model contains an Einstein-Hilbert
term and cosmological constant for the bulk as well as tension terms for the
branes
\begin{eqnarray}
S_{\rm RS} 
&=& M_5^3 \int d^4xdy\, \sqrt{-g}\, \left[ -2\Lambda + R \right]
\nonumber \\
&&
+ M_5^3 \int_{\rm UV} d^4x\, \sqrt{-h_0}\, \left[ -2\sigma_0 + 4K_0 \right]
\label{rsaction} \\
&& 
+ M_5^3 \int_{\rm IR} d^4x\, \sqrt{-h_1}\, \left[ -2\sigma_1 + 4K_1 +
M_5^{-3}{\cal L}_{\rm sm} \right] . 
\nonumber 
\end{eqnarray}
Here $h_{0,1}$ and $K_{0,1}$ are the determinant of the induced metric and the
trace of the extrinsic curvature on the UV and IR branes.  In terms of the AdS
curvature $k$, the cosmological constant is $\Lambda=-6k^2$ and the brane
tensions should be $\sigma_0=-\sigma_1=6k$.  ${\cal L}_{\rm sm}$ represents
the standard model Lagrangian.  The UV and IR branes are located at $y=0$ and
$y=\Delta y$ respectively.

To solve the strong CP problem in the low energy theory, we introduce a global
$U(1)_{\rm PQ}$ symmetry under which the brane quark and Higgs fields
transform non-trivially \cite{strongCP}.  Since the scale of Peccei-Quinn
symmetry breaking does not lie near the scales associated with either brane,
it is natural to attempt to break this symmetry through bulk dynamics.  Thus,
the axion will correspond to the phase of a bulk complex scalar field,
\begin{equation}
\sigma = {\rho\over\sqrt{2}} e^{ia} . 
\label{bulkscalar}
\end{equation}
The dynamics of this field will be determined by a $U(1)_{\rm PQ}$-symmetric
potential,
\begin{eqnarray}
S_\sigma 
&=& \int d^4xdy\, \sqrt{-g}\, \left[ - \nabla_a\sigma^\dagger \nabla^a\sigma -
V(\sigma^\dagger\sigma) \right] \nonumber \\
&&
+ \int_{\rm UV} d^4x\, \sqrt{-h_0}\, {\cal V}_0(\sigma^\dagger\sigma)
\nonumber \\
&&
+ \int_{\rm IR} d^4x\, \sqrt{-h_1}\, {\cal V}_1(\sigma^\dagger\sigma) , 
\label{PQaction}
\end{eqnarray}
Here, as in Goldberger and Wise \cite{gw}, the potentials on the branes will
be used to fix the value of the field $\rho$ on the UV and IR branes to be
respectively $\rho_0M_5^{3/2}$ and $\rho_1M_5^{3/2}$.  

The form of the $U(1)_{\rm PQ}$ symmetry breaking due to the bulk potential
$V(\rho)$ is quite different from the invisible axion solution in $3+1$
dimensions.  A vacuum solution in which $\rho=f$ for the bulk theory would
necessarily require some fine-tuning of the bulk potential to obtain a
realistic $f$.  Instead, $\rho$ can have some non-trivial dependence on the
extra dimension which also breaks the $U(1)_{\rm PQ}$.  In going to the low
energy effective theory, integrating out the bulk field $\rho$ will induce a
scale $f$ for the axion which plays the same role as the symmetry breaking
scale in the 4d invisible axion models.

The field equations for the components of $\sigma$ are 
\begin{eqnarray}
\nabla^2 a + {2\over\rho} \nabla_a\rho \nabla^a a &=& 0 \nonumber \\
\nabla^2\rho - {\delta V\over\delta\rho} - \rho \nabla_a a \nabla^a a &=& 0 . 
\label{eom}
\end{eqnarray}
The important dynamics of the axion occurs at energies well-below the TeV
scale beyond which bulk effects become important.  In this low energy regime,
we shall neglect the higher-order Kaluza-Klein modes of the axion which, since
it is a Goldstone mode, will have a massless mode which remains in the
effective theory.  Thus we shall consider only the lowest mode in the
Kaluza-Klein tower, $a\to a(x^\mu)$.  This situation differs greatly from a
bulk axion in models with large, flat extra dimensions where the Kaluza-Klein
modes of the axion are of the order of the inverse compactification radius and
are important in the low energy ($\ll$ TeV) theory \cite{ddg}.  The field
$\rho$ is not protected by any symmetry and its vacuum state is determined by
the bulk potential $V(\rho)$ so we shall neglect any $x^\mu$-dependent
fluctuations about the vacuum configuration, $\rho=\rho(y)$, as small in the
effective theory,
\begin{equation}
\rho^{\prime\prime} - 4k\rho' \approx {\delta V\over\delta\rho} 
\qquad\qquad
\partial_\mu\partial^\mu a \approx 0 . 
\label{decouple}
\end{equation}
The axion from this perspective becomes a massless field while the scalar
field $\rho$ has its dynamics set by the scale of the bulk physics.  At
energies below a TeV, we can integrate out $\rho(y)$ to obtain an effective
description of the axion dynamics,
\begin{eqnarray}
S_\sigma 
&=& \int d^4x\, \left[ - {1\over 2} 
\left( \int_0^{\Delta y}dy\, 2 e^{-2ky} \rho^2(y) \right)\partial_\mu a
\partial^\mu a \right]  \nonumber \\
&=& \int d^4x\, \left[ - {\textstyle{1\over 2}} f^2 \partial_\mu a
\partial^\mu a + \cdots \right] , 
\label{effaction}
\end{eqnarray}
where we have defined
\begin{equation}
f^2 \equiv \int_0^{\Delta y}dy\, 2 e^{-2ky} \rho^2(y)  
\label{fdef}
\end{equation}
which sets the scale associated with the axion by rescaling
\begin{equation}
a(x^\mu) \to {a(x^\mu)\over f} .
\label{axiondef}
\end{equation}
After this rescaling, the axion has the proper dimensions for a scalar field
in the 4d effective theory.

The remaining components needed to implement a solution to the the strong CP
problem closely resemble those found in standard invisible axion models.
Typically such models introduce heavy quarks which carry $U(1)_{\rm PQ}$
charge and couple to $\sigma$---KSVZ axions \cite{ksvz}---or an extra Higgs
doublet is added which couples to $\sigma$---DFSZ axions \cite{dfsz}.  To
obtain the latter model within a Randall-Sundrum scenario, we add an
interaction between a pair of brane Higgs doublets, $\Phi_1$ and $\Phi_2$, and
the bulk complex field $\sigma$,
\begin{equation}
S_{\rm int} = \int_{\rm IR} d^4x\, \sqrt{-h_1}\, \left[
\kappa M_5^{-1} \epsilon_{ij} \Phi_1^i \Phi_2^j (\sigma^\dagger(\Delta y))^2 
+ \hbox{h.c.} \right] ;
\label{dfszaction}
\end{equation}
here we have extracted a factor of the Planck mass so that $\kappa$ is a
dimensionless coupling.  Using that on the IR brane, $\sigma(\Delta y) =
{1\over\sqrt{2}}\rho_1 M_5^{3/2} e^{ia}$, and rescaling the axion using
Eq.~(\ref{axiondef}) and the Higgs fields by $\Phi_{1,2}\to e^{k\Delta
y}\Phi_{1,2}$ so that they have canonically normalized kinetic terms, the
leading behavior from Eq.~(\ref{dfszaction}) in the low energy limit is 
\begin{equation}
S_{\rm int} = \int d^4x\, \left[
\kappa_{\rm eff} \epsilon_{ij} \Phi_1^i \Phi_2^j e^{-2ia/f}  
+ \hbox{h.c.} \right] ,
\label{dfszeffact}
\end{equation}
where 
\begin{equation}
\kappa_{\rm eff} \equiv {\textstyle{1\over 2}} \kappa \rho_1^2 (e^{-k\Delta
y}M_5)^2 \sim {\cal O}({\rm TeV}^2) .
\label{kappadef}
\end{equation}

The standard model fields confined to the IR brane also have $U(1)_{\rm PQ}$
charges which we shall choose to be $+{1\over 2}$ for the right-handed
fermions and $-{1\over 2}$ for the left-handed $SU(2)$-doublet fermions.  With
these assignments, the Higgs fields have $U(1)_{\rm PQ}$ charges $+1$ so that
Eq.~(\ref{dfszeffact}) is an invariant interaction.  The fact that the Higgs
fields transform non-trivially under $U(1)_{\rm PQ}$ allows some of their
degrees of freedom to mix with the massless mode in the effective theory that
arises when we integrate out the extra dimension (\ref{effaction}).  At this
point the theory is essentially indistinguishable from 4d invisible axion
model.

\section{An intermediate brane}\label{intbrane}

The energy scale associated with the Standard Model fields remains naturally
light since they are confined to the IR brane at which the redshift suppresses
the strength gravity by an exponential factor.  Similarly, the introduction of
another brane, at some intermediate distance in the bulk, $0<y_a<\Delta y$,
will produce a new energy scale $e^{-ky_a}M_4$.  A simple mechanism for
achieving a reasonable value for the axion scale occurs when the bulk complex
scalar field is free with a mass of $m_\rho$.  If the brane potentials mainly
act to force $\rho$ to assume natural values on the intermediate and IR
branes, $\rho(y_a)=\rho_a M_5^{3/2}$ and $\rho(\Delta y) = \rho_1 M_5^{3/2}$
respectively with $\rho_a,\rho_1\sim {\cal O}(1)$, and to vanish on the UV
brane, $\rho(0)=0$, then integrating over the bulk yields a scale 
\begin{equation}
{f\over M_4} \approx {\sqrt{2}
(2+m_\rho^2k^{-2})^{1/4}\over(3+m_\rho^2k^{-2})^{1/2}} \rho_a e^{-ky_a} . 
\label{intbranef}
\end{equation}
Since we have assumed that the brane potentials in Eq.~(\ref{PQaction}) are
$U(1)_{\rm PQ}$ symmetric, as long as they are analytic functions of $\rho$,
$\rho=0$ will be an extremum on the branes so we do not need to fine-tune
$\rho(0)=0$ to be a minimum.  

To generate the scale for the axion, the relative positions of all three
branes must be stabilized.  In this section we shall show how the introduction
of a {\it single\/} real Goldberger-Wise \cite{gw} field stabilizes {\it
both\/} radion degrees of freedom that correspond to the two independent
distances between pairs of branes.  For simplicity we shall neglect the effect
of the bulk complex scalar $\rho$ when analyzing the brane stabilization.

Consider a bulk space-time in which the UV and IR branes reside as usual at
the fixed points of the orbifold, $y=0$ and $y=\Delta y$ respectively, while
an intermediate brane partitions the bulk into two regions with cosmological
constants $\Lambda_0=-6k_0^2$ ($0\le y\le y_a$) and $\Lambda_1=-6k_1^2$
($y_a\le y\le \Delta y$).  Matching the induced metric on both sides of the
axion brane, the bulk metric can be written in the form
\begin{equation}
ds^2 = e^{-2k_0y}\, \eta_{\mu\nu}\, dx^\mu dx^\nu + dy^2 
\label{IBbulka}
\end{equation}
for $0\le y\le y_a$ and 
\begin{equation}
ds^2 = e^{-2k_1y}e^{-2(k_0-k_1)y_a}\, \eta_{\mu\nu}\, dx^\mu dx^\nu + dy^2 
\label{IBbulkb}
\end{equation}
for $y_a\le y\le \Delta y$.  The Israel jump conditions across the branes
require the UV, IR and axion branes to have tensions respectively of
\begin{equation}
\sigma_0 = 6k_0 
\qquad
\sigma_1 = -6k_1 
\qquad
\sigma_a = 3(k_0-k_1) .  
\label{IBtensions}
\end{equation}
Note that when the cosmological constants are equal, $k_0=k_1=k$, the axion
brane becomes a tensionless `probe' brane.  For simplicity, we shall consider
this case in the following analysis.

Eq.~(\ref{IBtensions}) summarizes the three fine-tunings necessary for this
model.  One of these fine-tunings is equivalent to the vanishing of the
cosmological constant in the low energy effective theory.  As in the
Randall-Sundrum scenario, we shall not attempt to resolve this fine-tuning.
In a scenario with a further extra dimension, this vanishing can be reduced to
the tuning of the initial conditions rather than a tuning of the parameters in
the gravitational action \cite{warpRS}.  

The remaining two fine-tunings in Eq.~(\ref{IBtensions}) correspond to tuning
the two potentials for the positions of the IR and axion brane, relative to
the UV brane, to be flat.  The introduction of a bulk scalar produces an
effective potential, $V_{\rm eff}(y_a,\Delta y)$, which breaks both of the
symmetries associated with arbitrarily changing $y_a$ and $\Delta y$.  

Let us examine a single massive bulk scalar field,
\begin{equation}
S_\phi = M_5^3 \int d^5x\, \sqrt{-g}\, \left[ -{\textstyle{1\over
2}}\nabla_a\phi\nabla^a\phi - {\textstyle{1\over 2}} m^2 \phi^2 \right] 
\label{gwaction}
\end{equation}
with a mass $m = k\sqrt{\nu^2-4}$; we also define $\nu = 2+\epsilon$.  Note
that we have extracted a factor of $M_5^3$ so that $\phi(y)$ is dimensionless.
As in the standard Goldberger-Wise mechanism, we assume that the actions on
the three branes,
\begin{eqnarray}
S_\phi^{\rm brane} 
&=& M_5^3 \int_{y=0} d^4x\, \sqrt{-h}\, \left[ - \lambda_0 \left( \phi^2 -
v_0^2 \right)^2 \right] \nonumber \\
&& + M_5^3 \int_{y=y_a} d^4x\, \sqrt{-h}\, \left[ - \lambda_a \left( \phi^2 -
v_a^2 \right)^2 \right] \nonumber \\
&& + M_5^3 \int_{y=\Delta y} d^4x\, \sqrt{-h}\, \left[ - \lambda_1 \left(
\phi^2 - v_1^2 \right)^2 \right] \qquad
\label{gwbraction}
\end{eqnarray}
essentially act to fix the value of the scalar field to be $v_0$, $v_a$ and
$v_1$ on the UV, axion and IR branes respectively.  $h$ represents the
determinant of the induced metric on the appropriate brane.

The scalar field satisfies a Klein-Gordon equation in the bulk and its
solution in each of the two bulk regions is 
\begin{equation}
\phi(y) = \cases{\phi_0(y) &for $0\le y\le y_a$\cr
                 \phi_1(y) &for $y_a\le y\le \Delta y$\cr} 
\label{phicases}
\end{equation}
where
\begin{eqnarray}
\phi_0(y) &=&  
- {v_0  e^{-(\nu-2)ky_a} - v_a\over 1-e^{-2\nu ky_a}} e^{(\nu+2)k(y-y_a)} 
\nonumber \\
&&
+ {v_0 - v_a e^{-(\nu+2)ky_a}\over 1-e^{-2\nu ky_a}} e^{-(\nu-2)ky} 
\nonumber \\
\phi_1(y) &=& 
- {v_a e^{-(\nu-2)k(\Delta y-y_a)} - v_1\over 1-e^{-2\nu k(\Delta y-y_a)}} 
e^{(\nu+2)k(y-\Delta y)} 
\nonumber \\
&&
+ {v_a - v_1 e^{-(\nu+2)k(\Delta y-y_a)}\over 1-e^{-2\nu k(\Delta y-y_a)}} 
e^{-(\nu-2)k(y-y_a)} . \quad
\label{phisoln}
\end{eqnarray}
Integrating the scalar field action over the extra dimension produces an
effective potential for $y_a$ and $\Delta y$,
\begin{equation}
V_{\rm eff}(y_a,\Delta y) = M_5^3 \int dy\, e^{-4ky}\, \left[
\nabla_a\phi\nabla^a\phi + m^2 \phi^2 \right] ,
\label{effpotdef}
\end{equation}
which becomes, in terms of $z_a\equiv e^{-ky_a}$ and $z_1\equiv e^{-k\Delta
y}$,  
\begin{eqnarray}
{V_{\rm eff}(z_a,z_1)\over kM_5^3} &=& 
{(v_0 z_a^{\nu-2} - v_a)^2\over 1-z_a^{2\nu}} z_a^4 (\nu+2) \label{axpot} \\
&&
+ {(v_0 - v_a z_a^{\nu+2})^2\over 1-z_a^{2\nu}} (\nu-2) \nonumber \\
&&
+ {(v_a (z_1/z_a)^{\nu-2} - v_1)^2\over 1-(z_1/z_a)^{2\nu}} z_1^4 (\nu+2)
\nonumber \\
&&
+ {(v_a - v_1 (z_1/z_a)^{\nu+2})^2\over 1-(z_1/z_a)^{2\nu}} z_a^4 (\nu-2) .
\quad
\nonumber 
\end{eqnarray}
If this potential is to stabilize both of the radion parameters, then $z_a$
and $z_a$ are set by 
\begin{eqnarray}
{\partial V_{\rm eff}\over\partial\Delta y} &=& -kz_1{\partial V_{\rm
eff}\over\partial z_1} = 0 \nonumber \\ 
{\partial V_{\rm eff}\over\partial y_a} &=& -kz_a{\partial V_{\rm
eff}\over\partial z_a} = 0 .
\label{zeroderivs}
\end{eqnarray}
These first partial derivatives are 
\begin{widetext}
\begin{eqnarray}
{z_a\over kM_5^3} {\partial V_{\rm eff}\over\partial z_a} &=& 
{2z_a^4\over 1-z_a^{2\nu}} \biggl\{ 
(\nu^2-4) z_a^{\nu-2} \left[ v_0^2 z_a^{\nu-2} - 2v_0v_a + v_a^2 z_a^{\nu+2}
\right]  
+ 2 (\nu+2) \left[ v_0 z_a^{\nu-2} - v_a \right]^2 \label{sderiv} \\
&&\qquad\qquad
+ \nu(\nu+2) z_a^{2\nu} { \left[ v_0 z_a^{\nu-2} - v_a \right]^2\over
1-z_a^{2\nu}} 
+ \nu(\nu-2) z_a^{2(\nu-2)} { \left[ v_0 - v_a z_a^{\nu+2} \right]^2\over
1-z_a^{2\nu}} \biggr\} \nonumber \\
&&
- {z_1\over kM_5^3}  {\partial V_{\rm eff}\over\partial z_1} 
+ {4z_a^4\over 1-(z_1/z_a)^{2\nu}} \biggl\{ (\nu+2) \left( {z_1\over z_a}
\right)^4 \left[ v_a \left[ {z_1\over z_a} \right]^{\nu-2} - v_1 \right]^2  
+ (\nu-2) \left[ v_a - v_1 \left[ {z_1\over z_a} \right]^{\nu+2} \right]^2
\biggr\} 
\nonumber
\end{eqnarray}
and 
\begin{eqnarray}
z_1{\partial V_{\rm eff}\over\partial z_1} &=& 
{2z_1^4\over 1-(z_1/z_a)^{2\nu}} 
\biggl\{ (\nu^2-4) \left[ {z_1\over z_a} \right]^{\nu-2} \left[ v_a^2 \left[
{z_1\over z_a} \right]^{\nu-2} - 2v_av_1 + v_1^2 \left[ {z_1\over z_a}
\right]^{\nu+2} \right]  
+ 2 (\nu+2) \left[ v_a \left[ {z_1\over z_a} \right]^{\nu-2} - v_1 \right]^2
\nonumber \\
&&\qquad\qquad\qquad
+ \nu(\nu+2) \left[ {z_1\over z_a} \right]^{2\nu} { \left[ v_a
(z_1/z_a)^{\nu-2} - v_1 \right]^2\over 1-(z_1/z_a)^{2\nu}} 
+ \nu(\nu-2) \left[ {z_1\over z_a} \right]^{2(\nu-2)} { \left[ v_a - v_1
(z_1/z_a)^{\nu+2} \right]^2\over 1-(z_1/z_a)^{2\nu}} \biggr\} . \qquad
\label{rderiv}
\end{eqnarray}
\end{widetext}
Both $z_a$ and $z_1$ are exponentially small so higher powers of these factors
in Eqs.~(\ref{sderiv}--\ref{rderiv}) contribute negligibly.  The structure of
the matrix of second derivatives is such that as long as the intermediate
brane is not too close to the UV brane, one eigenvalue will be of the order
$z_a^4$ while the other will be of the order $z_1^4$.  This structure breaks
down when $z_a^3\sim z_1$ in which case the smaller eigenvalue receives
corrections of the order $z_a^{12}\sim z_1^4$.  For the scale required for the
axion, the intermediate brane will be sufficiently far from the UV brane that
we are well within the $z_1\gg z_a^3$ regime.  Retaining terms up to order
$z_1^4$, we have 
\begin{widetext}
\begin{eqnarray}
{z_a\over kM_5^3} {\partial V_{\rm eff}\over\partial z_a} &=& 
 4(2+\epsilon)z_a^4 \left[ 
(2+\epsilon) v_0^2z_a^{2\epsilon} - (4+\epsilon) v_0v_a z_a^\epsilon + 2v_a^2
\right] 
- 4\epsilon(2+\epsilon) r^4 v_a \left[ {z_1\over z_a} \right]^\epsilon \left[ 
v_a \left[ {z_1\over z_a} \right]^\epsilon - v_1
\right] + {\cal O}(z_a^8,z_1^8) .
\nonumber \\
{z_1\over kM_5^3} {\partial V_{\rm eff}\over\partial z_1} &=& 4 z_1^4  \left[ 
(2+\epsilon)^2 v_a^2 \left[ {z_1\over z_a} \right]^{2\epsilon} 
- (2+\epsilon)(4+\epsilon) v_av_1 \left[ {z_1\over z_a} \right]^\epsilon 
+ (4+\epsilon) v_1^2 
\right] + {\cal O}(z_1^8) 
\label{approxderivs} 
\end{eqnarray}
\end{widetext}
The second equation determines the equilibrium distance between the
intermediate and IR branes, 
\begin{equation}
\left[ {z_1\over z_a} \right]^\epsilon 
= {1\over 2(2+\epsilon) } {v_1\over v_a} 
\left[ (4+\epsilon) + \sqrt{ \epsilon(4+\epsilon)} \right] , 
\label{drzero}
\end{equation}
while the first equation, after using Eq.~(\ref{drzero}), determines the
distance between the intermediate and UV branes, 
\begin{equation}
z_a^\epsilon = 
{v_a\over v_0} 
\left[ 1 + {\sqrt{ \epsilon(4+\epsilon)} \over (2+\epsilon)^2} \left[
{z_1\over z_a} \right]^4 {v_1^2\over v_a^2} \right] .
\label{dszero}
\end{equation}
In Eqs.~(\ref{drzero}--\ref{dszero}) we have implicitly chosen the root which
corresponds to a minimum in both the $z_a$ and $z_1$ directions.  At these
extrema, the second partial derivatives are, to leading order in $\epsilon$
and in the small exponentials,
\begin{eqnarray}
{\partial^2 V_{\rm eff}\over\partial y_a^2} &=& 
8 \epsilon^2 v_a^2 M_4^2 e^{-4ky_a} 
\nonumber \\
{\partial^2 V_{\rm eff}\over\partial\Delta y^2} &=& 
16 \epsilon^{3/2} v_1^2 M_4^2 e^{-4k\Delta y}
\nonumber \\
{\partial^2 V_{\rm eff}\over\partial y_a\partial\Delta y} &=& 
- 16 \epsilon^{3/2} v_1^2 M_4^2 e^{-4k\Delta y} . 
\label{secdervs}
\end{eqnarray}
The eigenvalues of this matrix of second derivatives are both positive, 
\begin{equation}
8 \epsilon^2 v_a^2 M_4^2 e^{-4ky_a}
\quad\hbox{and}\quad
16 \epsilon^{3/2} v_1^2 M_4^2 e^{-4k\Delta y} , 
\end{equation}
so the relative positions of the branes are stable.

The bulk complex scalar which contains the axion might disrupt the
stabilization of the branes if it produces too large a contribution to the
total effective potential.  However, the contribution from the complex field
will be negligible compared to that arising from integrating out the field
$\phi$ as long as $v_0,v_a,v_1\gg \rho_a,\rho_1$.

\subsection*{Multiple branes}

In the preceding example with a single intermediate brane, we observed that
for each brane we have one fine-tuning.  Aside from the one fine-tuning for
the effective cosmological constant at low energies, these fine-tunings can be
interpreted as requiring the potentials for the radions---for the relative
separations among the branes---to vanish.  A single bulk scalar field was
sufficient to break the flatness with respect to both independent interbrane
distances.  This technique can be extended to scenarios with multiple
intermediate branes which could then have several large mass scales appearing
in the low energy theory.

As an example, consider a theory with two intermediate branes at positions
$y=y_a$ and $y=y_b$.  Including a free massive scalar field in the bulk with
potentials on these branes such that $\phi(y_a) = v_a$ and $\phi(y_b)=v_b$ but
otherwise identical to the case above, we find that the branes are stabilized
at 
\begin{equation}
ky_a = {1\over\epsilon} \ln\left[ {v_0\over v_a} \right] 
\quad\hbox{,}\quad
ky_b = {1\over\epsilon} \ln\left[ {v_0\over v_b} \right] 
\label{threebranesposa}
\end{equation}
and
\begin{equation}
k\Delta y 
= -{1\over\epsilon} \ln\left[ {1\over 2} {1\over 2+\epsilon} {v_1\over v_0} 
\left[ 4 + \epsilon + \sqrt{\epsilon(4+\epsilon)} \right] \right]
\label{threebranesposb}
\end{equation}
to leading order in powers of the small exponentials.  The matrix of second
derivatives for $y_a$, $y_b$ and $\Delta y$ given by
Eqs.~(\ref{threebranesposa}--\ref{threebranesposb}) has eigenvalues which are
all positive, 
\begin{eqnarray}
\lambda_a &\approx& 8\epsilon^2 v_a^2 M_4^2 e^{-4ky_a} \nonumber\\
\lambda_b &\approx& 8\epsilon^2 v_b^2 M_4^2 e^{-4ky_b} \nonumber\\
\lambda_1 &\approx& 16\epsilon^{3/2} v_1^2 M_4^2 e^{-4k\Delta y} , 
\label{threebranes}
\end{eqnarray}
to leading order in $\epsilon$.  In deriving Eq.~(\ref{threebranes}) we have
assumed that the branes are sufficiently separated, $e^{-ky_b}\gg e^{-3ky_a}$
and $e^{-k\Delta y}\gg e^{-3ky_b}$, for the eigenvalues to assume this form.

\section{Bulk potentials}\label{bulkpot}

While we can arrange for an invisible axion with the correct symmetry breaking
scale $f$ without any unnatural constraints on the theory, it becomes more
difficult to avoid some fine-tuning when we attempt to use a bulk potential to
produce this behavior.  In this section we shall determine how carefully we
need to tune the form of simple potentials for the bulk complex field to
produce a reasonable value for this scale.

As we saw for the inclusion of an intermediate brane, if $\rho\ll 1$ until
some intermediate position $y_a$, then the warping factor from the bulk metric
yields  $f\sim e^{-ky_a}M_4$.  We can obtain some intuition as to the
necessary form for the bulk potential by noting that the field equation for
$\rho$,
\begin{equation}
\rho^{\prime\prime} - 4k\rho' = {\delta V\over\delta\rho} , 
\label{rhoeqn}
\end{equation}
is that of a particle rolling in the inverted potential, $-V(\rho)$, under the
influence of a {\it negative\/} friction term.  If the particle starts at
$\rho=0$ with a small initial velocity, it tends to accelerate.  Thus, in
regions where the potential is approximately constant, its value will be
exponentially larger after a finite interval or, conversely, throughout most
of the interval $\rho(y)$ will be exponentially small.  This evolution should
not occur throughout the entire bulk since then the integral (\ref{fdef})
would then only produce an $f\sim e^{-k\Delta y}M_4 \sim {\rm TeV}$.  If
$V(\rho)$ decreases substantially after $\rho$ has become sufficiently large,
this change will act to dissipate the `kinetic energy' produced by the
friction term and $\rho$ will grow more slowly so that $\rho$ is not
exponentially weighted toward that latter end of this stage of its evolution
in the bulk.  After this dissipative stage, we could follow it with another
region in which $V(\rho)$ is approximately flat---as long as $\rho$ does not
grow exponentially larger than its values during the prior stage before it
reaches the IR brane, the exponential factor in Eq.~(\ref{fdef}) insures that
the integral will be dominated by intermediate values of $y$.

\subsection{A free massive bulk field}\label{mass}

The simplest potential is a mass term for the bulk scalar.  From the preceding
arguments, a positive mass squared term will have the effect of accelerating
the growth of the field already produced by the friction term as we move from
the UV to the IR brane.  Although it might seem that a negative mass term
could slow the effects of the friction term, we shall see that this case also
does not produce an acceptable value for $f$ without some fine-tuning.  Since
the field equations for this potential can be solved exactly, we present both
cases.

Consider a generalized mass term, to allow for either a stable or unstable
extremum at $\rho=0$,
\begin{equation}
V(\rho) = {\textstyle{1\over 2}}(\mu-4) k^2 \rho^2,
\label{masspotI}
\end{equation}
where $\mu$ is a dimensionless parameter.  When $\mu>0$, the solution is 
\begin{equation}
\rho(y) = \rho_1 M_5^{3/2} {e^{2ky}\over e^{2k\Delta y}}
{\sinh(\sqrt{\mu}ky)\over\sinh(\sqrt{\mu}k\Delta y)}
\label{posmu}
\end{equation}
while for $\mu<0$, 
\begin{equation}
\rho(y) = \rho_1 M_5^{3/2} {e^{2ky}\over e^{2k\Delta y}}
{\sin(\sqrt{-\mu}ky)\over\sin(\sqrt{-\mu}k\Delta y)}.  
\label{negmu}
\end{equation}
Here we have imposed the boundary conditions $\rho(0)=0$ and $\rho(\Delta y) =
\rho_1$.  For $\mu>0$, the scale $f$ is always of the order
\begin{equation}
f/M_4\sim \rho_1 e^{-k\Delta y}
\label{posmuf}
\end{equation}
which is too small for $\rho_1$ of a natural size, ${\cal O}(1)$.  In fact
$\rho_1$ should be somewhat small if the presence of the scalar field is not
to distort the background AdS$_5$ geometry.  For $\mu<0$ the zeros of the
denominator can generate much larger scales than (\ref{posmuf}); when
$\sqrt{-\mu}k\Delta y = n\pi-\delta$ ($n=1,2,3,\ldots$),
\begin{equation}
f/M_4\sim \rho_1 {\sqrt{-\mu}\over\sqrt{1-\mu}} {e^{-k\Delta y}\over\delta} ,
\label{negmuf}
\end{equation}
but only if we finely tune $\delta\alt 10^{-6}$.  

Note that in either case, if we relax the requirement $\rho(0)=\rho_0=0$ then
$\rho_0$ must itself be tuned to be of the order $\le 10^{-6}$.  In this case
$\mu$ should also satisfy $\mu>1$ since otherwise the large negative mass
squared favors an exponentially large value of $\rho(y)$ within the bulk so
that we would find $f\gg M_4$.

\subsection{A potential well}\label{potwell}

The reason that a mass term alone does not succeed is that the potential
contains no feature which might allow a brief growth of $\rho(y)$ which
appears in the vicinity of $0<y_a<\Delta y$ but which is damped soon after so
that the field assumes an ${\cal O}(1)$ value at the IR brane.

To model this behavior with a potential which we can solve exactly, we shall
study the following toy potential,
\begin{equation}
V(\rho(y)) = \cases{
V_0 &for $\rho\le\rho_a$ \cr
V_0\left[ 1 - {\rho(y)-\rho(y_a)\over \Delta\rho} \right] &for
$\rho_a\le\rho\le\rho_b$ \cr
0 &for $\rho\ge\rho_b$ \cr} 
\label{toypot}
\end{equation}
where
\begin{equation}
\Delta\rho = \rho_b - \rho_a . 
\label{Drhodef}
\end{equation}
The parameters specifying this potential are $\rho_a$, $\rho_b$ and $V_0$.  We
next shall estimate how carefully the form of the potential must be tuned to
achieve an acceptable value for $f$.

\begin{figure}[!tbp]
\includegraphics{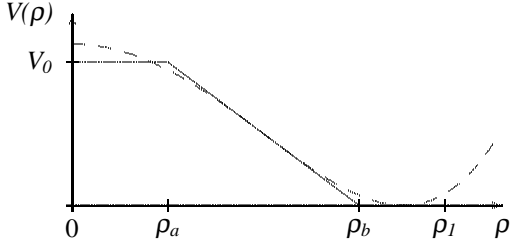}
\caption{We shall study the profile of a bulk field $\rho$ whose potential is
given by the toy model shown by the solid line.  It can be regarded as a crude
model for $0\le\rho\le \rho_1$ of a quartic, double-well potential shown by
the dashed line. 
\label{toypotfig}}
\end{figure}

Na\"\i vely, (\ref{toypot}) resembles a portion of a double well potential
seen in the vicinity of the origin, as shown in Fig.~\ref{toypotfig}, and we
assume that $V(-\rho)=V(\rho)$.  Note that $V(\rho)$ could grow again for
larger values of $\rho$, but as long as this growth occurs for values
$\rho(y)>\rho_1$, it will not affect our derivation.  Note also that we are
implicitly assuming that $\rho(y)$ is a monotonically increasing function of
$y$, which occurs provided the well is not so deep that all the `kinetic
energy' is dissipated and the particle rolls back toward $\rho=0$. 
It is also important that $\rho(0)=0$ at the UV brane, which as in the
intermediate brane case can be be arranged without any additional fine-tuning,
since it is difficult for any natural potential to suppress it quickly enough
to prevent the small $y$ region from dominating Eq.~(\ref{fdef}).  We shall
assume hereafter that $\rho_0=0$ and $\rho_1\approx{\cal O}(1)$.

Let us define positions $y_a<y_b$ such that $\rho(y_a)=\rho_a$ and
$\rho(y_b)=\rho_b$.  The solution to Eq.~(\ref{rhoeqn}) for this toy potential
with the boundary conditions $\rho(0)=0$ and $\rho(\Delta y)=\rho_1$ is then 
\begin{equation}
\rho(y) = c \left( e^{4ky} - 1 \right) 
\label{rhoone}
\end{equation}
for $0<y<y_a$,
\begin{eqnarray}
\rho(y) 
&=& c \left( e^{4ky} - 1 \right) - {1\over 16k^2} {V_0\over\Delta\rho} \left(
e^{4k(y-y_a)} - 1 \right) \nonumber \\
&&
+ {1\over 4k} {V_0\over\Delta\rho} (y-y_a) 
\label{rhotwo}
\end{eqnarray}
for $y_a<y<y_b$ and 
\begin{eqnarray}
\rho(y) 
&=& c \left( e^{4ky} - 1 \right) - {1\over 16k^2} {V_0\over\Delta\rho} \left(
e^{-4ky_a} - e^{-4ky_b} \right) e^{4ky} \nonumber \\
&&
+ {1\over 4k} {V_0\over\Delta\rho} (y_b-y_a) 
\label{rhotri}
\end{eqnarray}
for $y_b<y$.  For convenience we have defined the constant $c$ to be 
\begin{eqnarray}
c &\equiv& 
{\rho_1M_5^{3/2}\over e^{4k\Delta y} - 1}
+ {1\over 16k^2} {V_0\over\Delta\rho} \left( e^{-4ky_a} - e^{-4ky_b} \right) 
{e^{4k\Delta y}\over e^{4k\Delta y} - 1} \nonumber \\
&&
- {1\over 4k} {V_0\over\Delta\rho} {y_b-y_a\over e^{4k\Delta y} - 1} .
\label{cdef}
\end{eqnarray}
To learn whether the potential requires any fine-tunings, we can
reparameterize the slope of the potential in terms of the natural scales
available, $k$, $M_5$,  
\begin{equation}
{V_0\over\Delta\rho} \equiv 16k^2M_5^{3/2}\alpha 
\label{alphadef}
\end{equation}
where $\alpha$ should be some constant of order one.  To leading order in
powers of the exponential factors, the integral (\ref{fdef}) is then 
\begin{widetext}
\begin{eqnarray}
f^2/M_4^2 &=& 
{\textstyle{40\over 3}}\alpha^2 e^{-2ky_a} 
- {\textstyle{32\over 3}} \alpha^2 \left( 1+3k(y_b-y_a) \right) e^{-2ky_b} 
- {\textstyle{8\over 3}} \alpha^2 e^{-2k(y_b-y_a)} e^{-2ky_b}
\nonumber \\
&&
+ {\textstyle{1\over 3}} \left( \rho_1^2 + 16\rho_1k(y_b-y_a)\alpha
   - 128k^2 (y_b-y_a)^2\alpha^2 \right) e^{-2k\Delta y} 
\nonumber \\
&&
- {\textstyle{8\over 3}} \alpha \left( \rho_1 - 4k(y_b-y_a)\alpha \right) 
\left( 1 - e^{-2k(y_b-y_a)} \right) e^{-2k(\Delta y-y_b)} e^{-2k\Delta y}
+ \cdots  
\label{fullerff} 
\end{eqnarray}
\end{widetext}
For $e^{-ky_a}\gg e^{-ky_b}\gg e^{-k\Delta y}$ we have
\begin{equation}
{f\over M_4} \approx 2 {\textstyle\sqrt{10\over 3}} \alpha e^{-ky_a} . 
\label{ffortoy} 
\end{equation}
The chief contribution to (\ref{fdef}) comes from the region $y\sim y_a$.

We can now show that to achieve a realistic value for $y_a$ requires finely
tuning the potential.  In terms of the parameters $y_a,y_b$ of the solution we
have
\begin{eqnarray}
\rho_a M_5^{-3/2} &=& \alpha \left[ 1 - e^{-4k(y_b-y_a)} \right] + \cdots
\label{rhoabvals} \\
\rho_b M_5^{-3/2} &=& 4k\alpha(y_b-y_a) \nonumber \\
&& + e^{-4k(\Delta y - y_b)} \left[ \rho_1 - 4k\alpha(y_b-y_a) \right] +
\cdots . \nonumber
\end{eqnarray}
Note that $\rho_a$, $\rho_b$ and $\alpha$ are the parameters specifying the
shape of the potential.  The first line in Eq.~(\ref{rhoabvals}) indicates
that we must tune $\alpha^{-1}\rho_a M_5^{-3/2}\approx 1$ to within a
fractional correction of the order $e^{-4k(y_b-y_a)}$.  We can evade this
fine-tuning if $y_b\sim y_a$; however, then we must finely tune the value of
$\rho_b$ to within an order $e^{-4k(\Delta y-y_b)}$ correction.  From
Eq.~(\ref{ffortoy}) and Eq.~(\ref{fconstraint}), $16\alt ky_a \alt 23$ and an
electroweak-Planck hierarchy of $10^{-16}$ requires $k\Delta y \sim 37$.
Assuming $y_b\sim y_a$ yields then an exponentially small correction.

\section{Conclusions}\label{conclude}

The original Randall-Sundrum scenario contains only two, widely separated,
energy scales associated respectively with the bulk and the IR brane physics.
Since these scales could naturally be exponentially different, this model
provides an attractive alternative explanation of the hierarchy between the
gravitational and electroweak physics.  In this article we have shown that
other, intermediate scales can be incorporated into the Randall-Sundrum
scenario, such as are needed for the invisible axion solution to the strong CP
problem.

The basic requirement for generating a scale much lower than the Planck mass
from a bulk field is to find a mechanism which excludes this field from the
region near the UV brane.  The introduction of an intermediate brane produces
this behavior.  When the potential on the UV brane causes the field to vanish
there, most of the contribution to the effective symmetry breaking scale comes
from the region between the intermediate and the IR brane.  The position of
the intermediate brane then determines the scale needed by the invisible
axion.  In this scenario, the $U(1)_{\rm PQ}$ symmetry breaking proceeds
differently than in standard $3+1$ picture since it results from the
non-trivial profile of the bulk complex scalar field whose phase is the axion.

A single additional scalar field is needed to stabilize the two independent
distances between the pairs of branes.  From a low energy perspective, this
field produces an effective potential which breaks the necessity to tune the
potentials for the two radion parameters to be flat.  This stabilization only
requires a very mild tuning to ensure that the complex scalar does not disrupt
the mechanism.  The resulting radions have masses of the order of the
$U(1)_{\rm PQ}$ breaking scale and near the electroweak scale respectively.

While the invisible axion is not the only solution to the strong CP problem in
extra dimensional scenarios---for example the QCD gauge fields could be
promoted to bulk fields \cite{5dqcd,warp5dqcd}---it provides an intriguing
example of a warped geometry with multiple scales exponentially below the
Planck mass.  More generally, the existence of intermediate scales allows the
possibility of a hidden sector which is naturally suppressed by a large, but
not-Planckian, mass.  It would also be interesting to understand the origin of
multiple mass scales from the perspective of the AdS/CFT correspondence
\cite{malda}.

\begin{acknowledgments}
This work was supported in part by DOE grant DE-FG03-91-ER40682.
\end{acknowledgments}

\end{document}